\documentclass[aps,twocolumn]{revtex4}
\usepackage{graphicx}
\usepackage{color}
\usepackage{amsmath}
\begin{document}
\newcommand{\bea}{\begin{eqnarray}}
\newcommand{\eea}{\end{eqnarray}}
\newcommand{\be}{\begin{equation}}
\newcommand{\ee}{\end{equation}}
\newcommand{\bi}{\bibitem}
\newcommand{\D}{\mbox{$\Delta$}}
\newcommand{\mitD}{\mbox{${\mit \Delta}$}}
\renewcommand{\d}{\mbox{$\delta$}}
\renewcommand{\t}{\mbox{$\theta$}}
\renewcommand{\o}{\mbox{$\omega$}}
\renewcommand{\a}{\mbox{$\alpha$}}
\newcommand{\s}{\sigma}
\newcommand{\g}{\mbox{$\gamma$}}
\newcommand{\ep}{\mbox{$\epsilon$}}
\newcommand{\p}{\mbox{$\psi$}}
\newcommand{\bfk}{\mbox{${\bf k}$}}
\newcommand{\bfR}{\mbox{${{\bf R}}$}}
\newcommand{\bfr}{\mbox{${\bf r}$}}
\newcommand{\naR}{\mbox{$\nabla_{\bf R}$}}
\newcommand{\xipa}{\mbox{$\xi_{\parallel}$}}
\newcommand{\xipe}{\mbox{$\xi_{\perp}$}}
\newcommand{\xia}{\mbox{$\xi_{ani}$}}
\newcommand{\rr}{{\bf r}}
\newcommand{\kk}{{\bf k}}
\newcommand{\pp}{{\bf p}}
\newcommand{\qq}{{\bf q}}
\newcommand{\BSCCO}{{Bi$_2$Sr$_2$CaCu$_2$O$_8$ }}
\newcommand{\YBCO}{{YBa$_2$Cu$_3$O$_{7-\delta}$ }}
\newcommand{\nn}{\nonumber}
\newcommand{\G}{{\cal G}}
\newcommand{\bom}{{\bar\omega}}
\newcommand{\KK}{{\cal K} }
\newcommand{\EE}{{\cal E} }
\newcommand{\sgn}{{\rm sgn}}
\newcommand{\uG}{\underline{G}}
\newcommand{\ucG}{\underline{{\cal G}}}
\newcommand{\uGo}{\underline{G}^0}
\newcommand{\ug}{\underline{g}}
\newcommand{\uL}{\underline{L}}
\newcommand{\uM}{\underline{M}}
\newcommand{\uV}{\underline{V}}
\newcommand{\uVm}{\underline{V_1}}
\newcommand{\uU}{\underline{U}}
\newcommand{\uR}{\underline{R}}
\def\k{{\bf k}}
\def\r{{\bf r}}
\def\R{{\bf R}}
\def\p{{\bf p}}
\def\D{{\cal D}}


\title{Two impurities in a $d$-wave superconductor:
 local density of states}
\author{Lingyin Zhu$^1$, W.A. Atkinson$^{2
,1}$, and  P. J. Hirschfeld$^{1,3}$\\~}

\affiliation{$^1$Department of Physics, University of Florida,
Gainesville FL 32611\\$^2$Department of Physics, Southern Illinois
University, Carbondale IL 62901-4401\\\\$^{3}$Center for
Electronic Correlations and Magnetism, EP6, Univ.  Augsburg,
Augsburg Germany}
\date{\today}
%
%
\begin{abstract}

 We study the problem of two local potential scatterers in a $d$-wave
superconductor, and show how quasiparticle bound state wave
functions interfere.  Each single-impurity electron and hole
resonance energy  is in general split in the presence of a second
impurity into two, corresponding to one even parity and one odd
parity state.    We calculate the local density of states (LDOS),
and argue that scanning tunneling microscopy (STM) measurements of
2-impurity configurations should provide more robust information
about the  superconducting state than 1-impurity LDOS patterns,
and question whether  truly isolated impurities can ever be
observed.  In some configurations highly localized, long-lived
states are predicted. We discuss the effects of realistic band
structures, and how 2-impurity STM measurements could help
distinguish between current explanations of LDOS impurity spectra
in the BSCCO-2212 system.
\end{abstract}


\maketitle

\section{Introduction}
The study of isolated point-like impurities in $d$-wave
superconductors began with the observation that the scattered
wavepattern probed in tunneling experiments should inherit the
fourfold symmetry of the $d$-wave state\cite{scalapino}.  It was
pointed out shortly afterwards by Salkola {\em et al.}\cite{Balatsky},
following earlier work on the analogous $p$-wave problem by
Stamp\cite{Stamp}, that quasi-bound resonances should occur near
strongly scattering impurities.  Both the resonance energy and the
details of the spatial structure of the resonant wave-pattern depend
sensitively on electron correlations and a large body of theoretical
work exists exploring this relationship\cite{otherlocaldos}.  Assuming
that the important details of the scattering potential are understood,
careful studies of the local density of states (LDOS) near isolated
impurites provide a uniquely powerful probe of the superconducting
state.  Experimentally, the high temperature superconductors (HTSC),
notably Bi$_2$Sr$_2$CaCu$_2$O$_{8+\delta}$ (BSCCO), have been examined
with sensitive scanning tunneling microscopy (STM) techniques.  Images
of the local environment of impurities\cite{yazdani,davis} have
confirmed the existence of quasi-bound states near strongly scattering
impurities like Zn, but have led to new questions regarding the
microscopic model for impurities and the superconducting state itself.

Other types of inhomogeneities not directly correlated with
impurity resonances  have been discovered by STM measurements on
the cuprates. These include the observation of large,
quasi-bimodal nanoscale fluctuations of the superconducting order
parameter, together with spatially correlated variations of the
electronic structure and lifetime.\cite{davisinhom,davisinhom2}
More recently,
 the
observation of checkerboard patterns in the LDOS of vortex
cores\cite{Hoffman1} and in inhomogeneous samples in zero
field\cite{Kapitulnik} in BSCCO has led to speculation that
antiferromagnetic phases can be formed in regions where
superconductivity is supressed.\cite{vortexcore} A good deal of
subsequent theoretical work on the competition between $d$-wave
superconductivity and various exotic order parameters\cite{exotic}
has excited the high-$T_c$ community with the suggestion that STM
measurements could be revealing the presence of competing magnetic
or other exotic subdominant order, shedding light on the origins
of superconductivity itself.

 On the other hand, a  more conventional
but still fascinating explanation has been put forward by by
Hoffman et al.\cite{Hoffman2}, who argue that Friedel oscillations
related to quasiparticle scattering between nearly parallel
sections of the Fermi surface can lead to the observed
checkerboard patterns, which correspond to well-defined dynamical
peaks in momentum space. These general remarks were followed very
recently by a calculation by Wang and Lee\cite{WangDHLee} studying
 the Friedel oscillations of a single impurity in a $d$ wave
 superconductor in momentum space.  These
 calculations support the idea that what STM is seeing is primarily the
 effect of quasiparticle wavefunctions interfering with one another in a
 fluctuating disorder field  on a $d$-wave
 superconducting background.

 Studies of true {\it interference} of quasiparticle wave functions in the presence of
 more than a single impurity are
rare, however.  Numerical solutions of the
 many-impurity problem yield little insight into the mechanism of
 interference itself.  In principle, the problem of two impurities
 is the simplest one which displays the interference effects of
 interest for the present problem; we study it here in order to understand whether the
 STM analysis of isolated impurity resonances is indeed justified,
 and to test if the scenario proposed by Hoffman et
 al.\cite{Hoffman2} is tenable in the presence of
 interfering Friedel oscillations.
 There is a small amount of earlier work on two
impurities in a $d$-wave superconductor, most of it also
numerical. Onishi et al.\cite{Miyake} solved the Bogoliubov-de
Gennes (BdG) equations and presented a few local density of states
profiles for two impurities. In a work philosophically related to
ours, the interference of bound state wave functions in the fully
disordered system was discussed by Balatsky in the context of
supression of localization effects due to impurity band
formation\cite{balatskylocalization}. Micheluchi and Kampf have
recently exhibited numerically how impurity induced bound states
accumulate at low energies, and argued that the impurity band at
low energies could be studied from this
perspective.\cite{micheluchi} Finally, during the preparation of
this manuscript, a paper by Morr and Stavropoulos examined the
two-impurity problem with particular emphasis on predictions for
cuprate STM studies.\cite{Morr} We comment throughout the text on
comparisons with these previous works.


We begin in Section \ref{sec:1imp} by reviewing the solution to
the single impurity problem which has been studied by several
authors. In Section \ref{sec:2imp}, we set up the formalism for
the two-impurity problem and give the exact solution for the
$T$-matrix, as well as a simplified form for some special cases.
In Sec. \ref{sec:boundstate}, we discuss the dependence of the
resonance splittings on the orientation and magnitude of the
interimpurity separation $\bf R$. The resonant energy splittings
are argued to give important information which is qualitatively
different from that obtainable from 1-impurity resonance energies;
in particular, they allow one in principle to map out, by repeated
measurements of energies for impurity configurations at separation
$\R$, the spatial structure of the homogeneous superconducting
Green's function.
 Starting from two widely
separated impurities with $R\gg \xi_0$, where $\xi_0$ is the
coherence length, we then give analytical solutions for the
splittings which exhibit explicitly the  strong dependence on the
direction of $\bf R$ due to the $d$-wave nodes. Impurities located
along a 100 axis  interact very weakly for $R \gtrsim\xi_0$,
whereas configurations near 110 relative orientation lead to
strong hybridization.

In Section \ref{sec:ldos}, we begin by discussing the simple
quantum mechanics problem of how bound eigenstates with two
scattering centers are constructed from the eigenstates of one,
and show how these states may be classified.  Of four states which
arise from the single impurity particle and hole resonances, two
are spatially symmetric (s), and two antisymmetric (p), one on
each side of the Fermi level.  The energy ordering of these
states, as well as the crude qualitative interference pattern
(constructive or destructive)
 depends on the configuration $\bf R$ in a
predictable way.    The local density of states (LDOS) is then
calculated and plotted for several cases to illustrate the spatial
dependence of the resonant state wave functions.  Spectra on
individual sites reveal unexpected phenomena: in certain
circumstances the 1-impurity states, which become sharper as they
approach the Fermi level due to the coupling to the linear
$d$-wave continuum, interfere to create localized, extremely sharp
states located at energies quite far from the Fermi level. In
Section \ref{sec:realistic} we discuss the situation for a more
realistic band characteristic of the BSCCO-2212 system on which
impurity STM studies have been performed.  The trapped
quasiparticle states are still found, and a dynamical resonanance
criterion depending on the exact Fermi surface and impurity
orientation $\R$ is identified.

Finally, in section VII we present our conclusions and discuss the
implications for STM experiments and other aspects of the
disordered quasiparticle problem.

\section{Single impurity problem}
\label{sec:1imp} In broad terms, there are two distinct points of
view which have evolved regarding the localized scattering
resonances observed by STM on the surface of BSCCO-2212. In the
traditional quasiparticle picture, one simply calculates the
scattering T-matrix for non-interacting BCS quasiparticles in a
$d$-wave superconductor and finds a pair of resonances which are
approximately symmetric in energy about the Fermi level, and which
have distinctive spatial patterns\cite{otherlocaldos}.  In the
second point of view, additional physics arises from the local
disruption of the strongly-correlated ground state by the impurity
which is predicted to break singlet correlations\cite{vojta,park},
nucleate short-ranged antiferromagnetic
order\cite{Khalliulin,Dagotto,Tanaka,Ting,Liang,Lee}, or spin
fluctuations associated with a nearby phase
transition\cite{Morr,vojtacollective}. Perhaps the strongest
motivation for this point of view is the observation of local
moments near Li and Zn impurities in the superconducting state of
underdoped YBCO, as observed in NMR\cite{alloul,Julien}
experiments.  For the STM experiments, the basic question is
whether the spatial structure which is observed at low energies is
effectively the result of BCS quasiparticles scattering from a
short-range potential (ie.\ Friedel oscillations), or requires
consideration of local correlation effects (eg.\ spin density wave
formation\cite{vojtacollective}) or dynamical effects (Kondo
physics\cite{Ingersent,vojta}).

One notable feature of the STM experiments on Zn-doped samples of
BSCCO is that only a single, negative energy resonance with a
resonance energy of $\approx -1.5$ meV is seen;\cite{davis} the
predicted impurity-induced resonances come in mirror pairs as a
consequence of the particle-hole symmetry of the superconducting
state.  A further inconsistency is the fact that a large LDOS is
observed on the Zn impurity site; the strong repulsive potential
which Zn is believed to possess must necessarily allow little or
no electron spectral weight at the impurity site. One appealing
explanation\cite{Zhufilter,martinbalatsky} is that the measured
LDOS of the CuO$_2$ planes is in fact {\em filtered} by an inert
surface layer, leading to an apparent redistribution of spectral
weight in the tunneling LDOS.  With this mechanism, one can simply
understand the LDOS without introducing strong correlation
physics.  We note, however, that while this mechanism explains the
observed LDOS for Zn impurities, it is problematic for both Ni
impurities and Cu vacancies, which are consistent with the
quasiparticle picture without invoking a filtering mechanism.  (To
date, there is no convincing model which has explained the spatial
distribution of the LDOS in all three cases.) One of the goals of
this work is to study the effect of the filtering mechanism on the
resonant structure of two closely-spaced strongly-scattering
impurities within the quasiparticle point of view, providing a
more rigorous test of the quasiparticle-plus-filter mechanism for
Zn impurities.

The BCS Hamiltonian for a pure singlet
superconductor can be written as:
\begin{equation}
H_{0}=\sum_{\k}\Phi_{{\k
}}^{\dagger}(\epsilon_{\k}\tau_{3}+\Delta_{k}\tau_{1})\Phi_{\k},
\end{equation}
where $\Phi_{\k}=(c_{\k\downarrow} c^{\dagger}_{-\k\uparrow} )$, is a
Nambu spinor.  In this work we will consider several forms for the
dispersion $\epsilon_\k$. Analytic results are presented for a
parabolic band $\epsilon_\k =k^2/2m$, with corresponding $d$-wave
order parameter $\Delta_\k=\Delta_{max}\cos 2\phi$ ($\phi$ is the
angle in momentum space which ${\bf k}$ makes with the 100 axis).
Numerical results are presented for a simple tight binding model
$\epsilon_\k=-2t(\cos k_x +\cos k_y) -\mu$ and for a realistic
6-parameter tight-binding model proposed by Norman {\em et
al.}\cite{Norman}, both having the corresponding $d$-wave order parameter
$\Delta_\k= \Delta_0(\cos k_x -\cos k_y)$.  Note the maximum value of
the order parameter in the lattice system with the current convention
is $2\Delta_0$.  The matrices $\tau_{i}$ are the Pauli matrices.

  The Hamiltonian of a single on-site impurity at $\r=0$ may be
written as $H_{imp}=\sum_{\k,{\k^{'}}}V_0
\Phi_{\k}^{\dagger}\tau_3\Phi_{{\k^{'}}}$,where $V_0$ is the strength
of the impurity potential.  The Green's function
$\hat{G}_{\k{\k^{'}}}(\omega)$ in the presence of the impurity is
expressed in terms of the Green's function $\hat{G}_{\k}^{0}(\omega)$
for the pure system as $\hat{G}(\k,\k^{'},\omega)=
\hat{G}^{0}(\k,\omega)\delta_{\k{\k^{'}}}+\hat{G}^{0}(\k,\omega)
\hat{T}(\omega)\hat{G}^{0}(\k^{'},\omega)$, where the $\hat{~}$ symbol
indicates a matrix in Nambu space. The solution is
\begin{eqnarray}
    {\hat T}&=&T_0\tau_0+T_3\tau_3\nonumber\\
    T_0&=&V_0^2G_0/(S_+S_-)\nn\\
    T_3&=&V_0^2(c-G_3)/(S_+S_-),
    \label{1imptmat}
\end{eqnarray}
where $G_0$ and $G_3$ are the $\tau_0$ and $\tau_3$ Nambu components
of the integrated bare Green's function $\sum_\k {\hat G}^0(\k,\omega)$.
This expression has resonances when
\begin{equation}
    S_\pm\equiv 1-V_0(G_3 \mp G_0)=0.
    \label{1impresonance}
\end{equation}
Note that in a simple band $1/(\pi N_0 V_0)\equiv c/(\pi N_0)$ is the
cotangent of the $s$-wave scattering phase shift $\eta_0$, where $N_0$
is the density of states at the Fermi level. In the special case of a
particle-hole symmetric system, $G_3$=0 and the resonance energy is
determined entirely by $G_0$, which is given in the case of a circular
Fermi surface by $G_{0}(\omega)=-i\int
\frac{d\varphi}{2\pi}\omega[\omega^2-\Delta_\k^2]^{-1/2}$ which for
low energies $\omega \ll \Delta_{max}$ takes the form
$G_0(\omega)\simeq -(\pi\omega/\Delta_{max})(\log 4\Delta_{max}/\omega
+i)$. One may then solve Re $S_\pm(\omega+i0^+)=0$ and estimate the
resonance width $\Gamma$ on the real axis.  In the case of strong
scattering $c\ll N_0$, the resonance energy $\Omega_0^\pm$ and
scattering rate $\Gamma$ are
\begin{subequations}
\begin{eqnarray}
     \Omega_0^\pm &=& {\pm \pi c\Delta_0\over 2 \log (8/\pi c) }
     \label{1impresenphsA} \\
     \Gamma &=& {\pi^2 c\Delta_0\over 4 \log^2 (8/\pi c)} .
     \label{1impresenphsB}
\end{eqnarray}
\end{subequations}
This result was first obtained by Balatsky et al.\cite{Balatsky}, following
earlier work on the $p$-wave analog problem by Stamp\cite{Stamp}. Note that the
resonance becomes a true bound state only exactly at the Fermi level
$\Omega=0$, when $c=0$; for finite $c$ there are two  resonances whose energies
are symmetric,  $\Omega_0^+=-\Omega_0^-$ in this approximation. As seen from
(\ref{1impresonance}), in particle-hole asymmetric systems ($G_3\ne 0$), the
resonance is tuned to sit at the Fermi level for some value of the impurity
potential $V_0$ which is not infinite, so the term ``unitarity" (as used in
this work, $\Omega_0=0$) and ``strong potential" ($V_0\rightarrow \infty$) are
not synonymous.\cite{joynt,ahmphysica}

The ambiguity in defining the resonance energy precisely arises
already at the level of the 1-impurity problem.  Eq.
(\ref{1impresonance}) is in fact an equation for a complex
frequency $\omega$, which may be shown to have no solution in the
upper half plane. Thus the $T$-matrix has no true pole at any
$\omega=\Omega^\prime+i\Omega^{\prime\prime}$ in the complex
plane, but only a maximum which lies along the real axis. Only
when the real part $\Omega'$ approaches the Fermi level does the
damping become sufficiently small to allow one to speak of a
well-defined resonance. in this case the real part $\Omega'$
approaches the solution $\Omega_0^\pm$ of $Re[S_\pm]=0$ for
$\omega$ on the real axis.  For this reason the ``resonance
energy" is usually taken to be $\Omega_0^\pm$, as given, e.g. in
Eq. (\ref{1impresenphsA}) for the particle-hole symmetric case.
It is important to keep in mind, however, that the definition
becomes meaningless as the bound state energy moves far from the
Fermi level; for example, the apparent divergence of
(\ref{1impresenphsA}) as $c\rightarrow 8/\pi$ is artificial, since
no well-defined resonant state exists by the time $\Omega_0$ is a
significant fraction of the gap $\Delta_0$.  It should also be
noted that the generalization of this resonance criterion to more
complicated situations, where the denominator does not factor, is
not straightforward.

\begin{figure}
\begin{center}
\leavevmode
\includegraphics[width=\columnwidth]{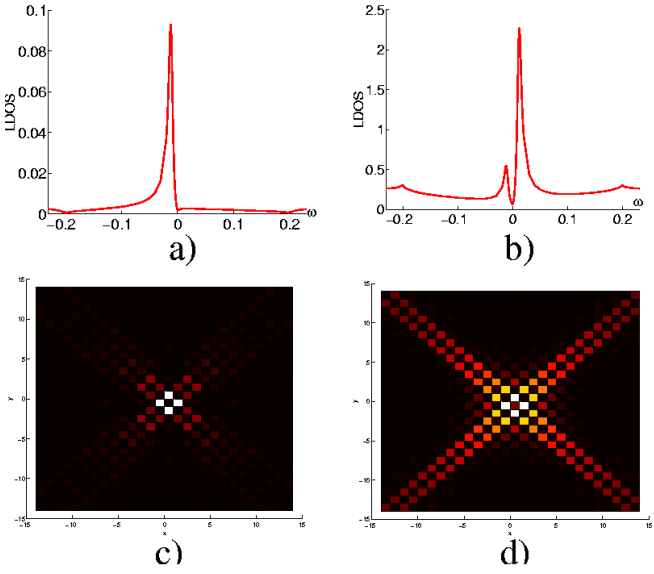}
\caption{Summary of LDOS results for 1-impurity problem on a
tight-binding lattice, $\Delta_0=0.1, \mu=0, V_0=10$,
$|\Omega_0^\pm|\simeq 0.013$: a) LDOS vs. $\omega$ on the impurity
site; b) LDOS vs. $\omega$ on nearest neighbor site; c),d) LDOS
map at resonance $\omega=\Omega_0^+$ and $\omega=\Omega_0^-$.
Color scales in c) and d) are relative to the nearest neighbor
peak heights shown in b).} \label{fig:1imp}
\end{center}
\end{figure}

 Even if the {\it energies} of the particle and hole resonant states
are symmetric, their {\it spectral weights} on a given site may be
quite different.\cite{Balatsky,otherlocaldos} The finite impurity
potential acts as a local breaker of particle-hole symmetry,
leading in the case of repulsive potential to a large peak in
$\rho(\r,\omega)$ at the impurity site $\r=0$ at negative
energy(holelike states) and a small feature at positive energy
(electronlike), as seen in Fig.\ \ref{fig:1imp}. The
impurity-induced LDOS decays as $r^{-2}$ along the nodal
directions (for the particle-hole symmetric system), and
exponentially along the antinodes. The LDOS in the near field is
more complicated, however: the nearest neighbour sites have peaks
at $\pm\Omega_0$, with the larger spectral weight at $+\Omega_0$.
In the crossover regime $r\sim \xi_{0}$, the LDOS is enhanced
along the node direction for holelike states, but is spread
perpendicular to the node direction for electronlike states. These
spatially extended LDOS patterns are the fingerprint of the
impurity-induced virtual bound states. In Figure \ref{fig:1imp},
we illustrate the LDOS pattern expected for both particles and
holes for a resonant state close to the Fermi level. Results are
obtained with a simple half-filled tight-binding band,
$\epsilon_\k=-2t(\cos k_x+\cos k_y)$, and unless otherwise
specified all energies are given in units of the hopping $t$.

At the present writing, the LDOS pattern produced within the
simple $T$-matrix  theory for a single strong impurity, while
4-fold in symmetry, does not agree in detail with STM experiments
on Zn impurities and native planar defects.\cite{davis}  It is not
currently clear whether this is due to a failing of the
microscopic impurity model, e.g. failure to include strong
correlations in the host or magnetic degrees of freedom or whether
a relatively trivial tunneling matrix element effect prevents
direct observation of the simple pattern by
STM\cite{Zhufilter,martinbalatsky}. For the moment then we
consider only the simplest 2-impurity model possible, recognizing
that direct application to experiments awaits a resolution of the
discrepancy at the 1-impurity level.

\section{T-matrix for 2 impurities}
\label{sec:2imp} We now introduce the formalism necessary to study
the interference between two resonances of the type discussed
above when two identical impurities are brought close to one
another. The perturbation due to 2 identical impurities located at
positions $\R_\ell$ and $\R_m$, is given by
\begin{equation}
\hat H_{imp} = V_0 \sum_{i=\ell,m} {\bf \Phi}_i^\dagger \tau_3
    {\bf \Phi}_i
\end{equation}
where ${\bf \Phi}_i \equiv [c^\dagger_{i\uparrow} c_{i\downarrow}]$. By
iterating the procedure for the single impurity T-matrix with two
impurities present we find, in a $4 \times 4$ basis of spin and impurity site
labels:\cite{economou}
\begin{equation}
    \hat T_{\ell m}(\omega)=\left ( \begin{array}{cc}
    \hat f\hat T_\ell &
    \hat f\hat T_\ell \hat G^{0}(\R)\hat T_m\\
    \hat f\hat T_m\hat G^{0}(-\R)\hat T_\ell&
    \hat f\hat T_m
    \end{array} \right )
\label{2impT}
\end{equation}
where $\R = \R_\ell - \R_m$ and where $\hat T_{\ell},\hat T_{m}$ are
the single impurity T-matrices associated with the two impurities.
For identical impurities, $\hat T_\ell = \hat T_m = \hat T(\omega)$,
the single impurity T-matrix defined previously.  The quantity $\hat
f$ is defined as:
\begin{equation}
    \hat f(\omega) =[1-\hat G^{0}(-\R,\omega) \hat T_\ell(\omega)
    \hat G^{0}(\R,\omega)\hat T_m(\omega)]^{-1},
\end{equation}
where $\hat G^{0}(\R,\omega) = \sum_\k
\exp[i\k\cdot\R]G^0_\k(\omega)$ is just the Fourier transformation
of $\hat G^{0}_\k(\omega)$, the unperturbed Nambu Green's
function. For systems with inversion symmetry $\hat G^0(\R,\omega)
= \hat G^0(-\R,\omega)$.
Note that in Eq. (\ref{2impT}), the physical processes are clearly
identifiable as multiple scatterings from each impurity $\ell$ and
$m$ individually, plus interference terms where electrons scatter
many times between $\ell$ and $m$.
In $\k$-space, we can write the T-matrix in the more
usual $2 \times 2$ notation as
\begin{eqnarray}
%
    \hat T_{\k\k^\prime}(\omega) =
    [ e^{i\k\cdot \R_l}\tau_0 &e^{i\k\cdot \R_m}\tau_0 ]
    \hat T_{\ell m}
    \left [ \begin{array}{c}
     e^{-i\k^\prime \cdot \R_l}\tau_0  \\
    e^{-i\k^\prime \cdot \R_m}\tau_0
    \end{array} \right ]
\end{eqnarray}
where $\tau_0$ is the Pauli matrix.
In Sec.\ \ref{sec:1imp} we showed that, provided the resonance
energies are distinct, peaks in the total density of states
correspond to minima of the T-matrix denominator:
\begin{equation}
    {\cal D}\equiv {\rm det}[1-\hat G^{0}(-\R,\omega) \hat T(\omega)
    \hat G^{0}(\R,\omega)\hat T(\omega)].
    \label{2impexplicitdenom}
\end{equation}
Explicitly, ${\cal D}={\cal D}_{1}{\cal
D}_{2}/(S_{+}^{2}S_{-}^{2})$ with
\begin{eqnarray}
{\cal D}_{1}&=& {\cal D}_1^+ {\cal D}_1^-+V_0^2G_{1}^{2}({\R},\omega)\nn\\
{\cal D}_{2}&=&{\cal D}_2^+ {\cal
D}_2^-+V_0^2G_{1}^{2}({\R},\omega)\label{d1d2}
\end{eqnarray}
where
\begin{eqnarray}
{\cal D}_{\alpha}^\pm&=&[1-V_{0}G_{3}(0,\omega)\pm V_{0}G_{0}(0,\omega)]\nonumber\\
&&+(-1)^\alpha V_{0}[\mp G_{0}(\R,\omega) + G_{3}(\R,\omega)] .\nn\\
\label{2impfactorizedD}
\end{eqnarray}


 The factors ${\cal D}_{1},{\cal{D}}_{2}$  determine the four
2-impurity resonant energies.
%
%
Here $G_{\alpha}(\R,\omega)$ is the $\tau_{\alpha}$ component of the integrated
bare Green's function
\begin{equation}
G_\alpha (\R,\omega)= {1\over 2} {\rm Tr}\left(\tau_\alpha G^0(\R,\omega)
    \right).
\end{equation}

For completeness, we also give the explicit form of the $T$-matrix itself: \bea
{\hat T}_{\k,\k'}(\omega) &=&{2V_0\over{\cal D}_1{\cal D}_2}\left( \cos
(\k\cdot \R /2)\cos(\k'\cdot\R/2) \,{\hat M}_s(\omega) \right.\nn\\&& +
\left.\sin( \k\cdot \R/2) \sin(\k'\cdot\R/2) \,{\hat M}_p (\omega)\right),
\label{explictT} \eea where
%
%
%
\begin{eqnarray}
{\hat M}_{p}&=&\left(\begin{array}{cc} {\cal D}_{2}^{+}{\cal
D}_{1} &
G_{1}(R,\omega)V_0{\cal D}_{1}\\
G_{1}(R,\omega)V_0{\cal D}_{1} & -{\cal D}_{2}^{-}{\cal D}_{1}
\end{array} \right)\nonumber\\
{\hat M}_{s}&=&\left(\begin{array}{cc} {\cal D}_{1}^{+}{\cal D}_{2}& -G_{1}(R,\omega)V_0{\cal D}_{2}\\
-G_{1}(R,\omega)V_0{\cal D}_{2}& -{\cal D}_{1}^{-}{\cal D}_{2}
\end{array} \right).
\end{eqnarray}

%
%
%
 \vskip .2cm
In certain special configurations, e.g. if the two impurities are located at
45$^\circ$ with respect to one other, it is easy to check that
$G_1(\R,\omega)=0 $ $\forall$ $\R$.  In this case the entire resonant
denominator factorizes ${\cal D}={\cal D}_{1+} {\cal D}_{1-} {\cal D}_{2+}
{\cal D}_{2-}$. The $T$-matrix then takes the simple diagonal form

\bea {\hat T}_{\k,\k'}(\omega) &=& 2V_0 \cos (\k\cdot {\R\over 2}
)\cos(\k'\cdot{\R\over 2})\left[ {\tau_+\over {\cal D}_{1-}}+{\tau_-\over {\cal
D}_{1+}} \right] \nn
\\&+&2V_0 \sin (\k\cdot {\R\over 2})\sin(\k'\cdot{\R\over 2})\left[ {\tau_+\over {\cal
D}_{2-}}+{\tau_-\over {\cal D}_{2+}} \right],\nn\\&&~~~ \label{simpleT}\eea
where $\tau_\pm\equiv (\tau_3\pm \tau_0)/2$.

\section{Bound state energies for two impurities}
\label{sec:boundstate}

%
%

Measuring bound state energies of impurity resonances in STM
experiments allows one to obtain information on impurity
potentials, and has the virtue of being independent of the STM
tunnelling matrix elements.  On the other hand, resonance energies
of isolated single impurities provide no information on the
spatial structure of resonant or extended state electronic
wavefunctions.  In principle, measurement of only the resonance
energies of isolated pairs of impurities with different
separations $\R$ is the simplest method of getting spatially
resolved information on electronic wave functions independent of
the exact tunnelling mechanism.

When two identical impurities with resonance energies
$\Omega_0^-,\Omega_0^+$ are brought together, the bound state
wavefunctions interfere with one another, in general splitting and
shifting each resonance, leading to four resonant frequencies
$\Omega_1^-,\Omega_1^+,\Omega_2^-$ and $\Omega_2^+$, where the subscript
indicates which factor in Eq.\  (\ref{2impfactorizedD}) is resonant.
If splittings are not too large, the electron and hole resonances are
related in a similar way as in the 1-impurity problem,
$\Omega_1^-\simeq-\Omega_1^+$ and $\Omega_2^-\simeq-\Omega_2^+$. Again
the weight of each resonance may be quite different or even zero on
any given site.  A large splitting may be taken as evidence for strong
hybridization of quasiparticle wavefunctions. If we take the
interimpurity distance $R$ as a parameter and keep impurity potentials
and other parameters fixed, there are two obvious limits where this
splitting vanishes.  In the case of separation $R=0$, the two
impurities combine (mathematically) to create a single impurity of
strength $2V_0$, so both $\Omega_{1,2}^\pm$ approach the
$\Omega_0^\pm(2V_0)$ appropriate for the double strength potential.
In the case of infinite separation $R\rightarrow \infty$, we must find
$\Omega_{1,2}^+$ approaching the $\Omega_0(V_0)$ appropriate for
isolated single impurities.

\subsection{Gas model}
\label{subsec:gasmodel}  Eq. (\ref{2impexplicitdenom}) is a
general result for two $\delta$-function potentials embedded in a
host   described by an arbitrary $G_0$.  We would like to derive
analytical results for the resonance energies obtained therefrom
to get some sense of the appropriate length scales and symmetries
in the problem. At large distances, the resonance energies must
approach the single impurity values, so the splittings can be
calculated perturbatively.  To do so one must first obtain
analytical expressions for the large-distance behavior of the
unperturbed Green's functions.  This is difficult for the
superconducting lattice tight-binding model on which most of this
work is based, but much insight can be gained by studying the
equivalent gas model, with spectrum $\epsilon_\k=\k^2/2m$. In this
case expressions have been obtained  by Joynt\cite{joynt} and
Balatsky et al.\cite{balatskylocalization} for the $d$-wave
integrated Green's functions $G_\alpha(\R,\omega)$ at large
distances, both for $\R$ making an angle $45^\circ$ or $0 ^\circ$
with the $x$ axis.  For frequencies $\omega/\Delta_0\ll1/k_F r\ll
1/k_F\xi_0$, these reduce to
\begin{eqnarray}
{\hat G}^0(\R,\omega)
~~~~~~~~~~~~~~~~~~~~~~~~~~~~~~~~~~~~~~~~~~~~~~~~~~~~~~~~~~~~\\
\approx
\left\{
\begin{array}{cc} N_0{e^{ik_FR}\over k_FR} {k_F\xi_0\over 4+\pi^2\xi_0^2k_F^2}\tau_3
& {\bf R}\parallel (110)\\
N_0 {e^{-R/\xi_0}\over \sqrt{k_FR}} \left[ (i{\omega\over
\Delta_{max}} \tau_0 + \tau_1 +\tau_3 )\cos k_FR +\right.
&{\bf R}\parallel (100)\\\left. (i{\omega\over \Delta_{max}} \tau_0 +
\tau_1 -\tau_3 )\sin k_FR\right] & ~
\end{array}
\right.\nn \label{GR}
\end{eqnarray}
where $\xi_0=v_F/\pi\Delta_{max}$ is the coherence length.

The resonance energies may now be found by inserting these
expressions for frequencies $\omega=\Omega_0^\pm+\delta$ into
(\ref{2impexplicitdenom}) and solving for the shifts $\delta$.  We
find $\Omega_{1,2}^+\simeq \Omega_0^+\pm\delta$, with
\begin{eqnarray}\label{splittings}~~\\
{\delta\over \Delta_{max}}\approx\left\{
\begin{array}{cc}
{1\over\log (\Omega_0/\Delta_{max}) }{\sin{k_F R}\over k_F R}
{k_F\xi_0\over 4+\pi^2\xi_0^2k_F^2}
 & \R\parallel (110)\\
{e^{-R/\xi_0}\over \sqrt{k_FR}\log
(\Omega_0/\Delta_{max})}\cos(k_FR+\pi/4)&\R\parallel (100)
\end{array}
\right.\nn
\end{eqnarray}
These expressions are valid for $\delta/\Omega_0^+ \ll 1$.

Clearly the decay of the splitting $\sim \exp
-r/\xi_0/\sqrt{k_Fr}$ is much more rapid for distances larger than
the coherence length along the antinode (100) than for along the
nodes, where it falls as $\sim 1/r$.  The lack of a scale in the
long-distance interference of quasiparticle wavefunctions oriented
along (110), where they strongly overlap, is of potentially
crucial importance in the STM analysis of ``isolated" impurities,
and we will bear this question in mind in what follows.

\subsection{Lattice model}
\begin{figure}[tb]
\begin{center}
\leavevmode
\includegraphics[width=.9\columnwidth]{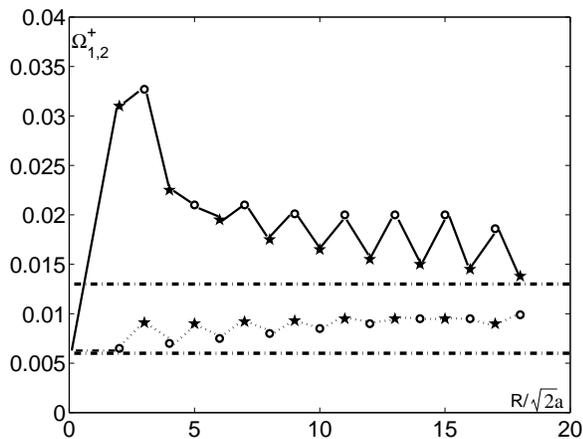}
\caption{2-impurity resonance energies $\Omega^+_{1,2}>0$ vs.
impurity separation $R/(\sqrt{2}a)$ for $\R\parallel (110)$,
$\mu$=0, $\Delta_0=0.1$, $V_0=10$.  Solid line: upper 2-impurity
positive resonance energy;
 dashed line: lower 2-impurity resonance energy; upper
dashed-dotted line: reference 1-impurity  resonance energy
$\Omega_0^+(V_0)$. Lower dashed-dotted line: 1-impurity resonance
energy for double impurity strength $\Omega_0^+(2V_0)$.  Symbol
indicates resonant channel: ${\circ}$=$\Omega^+_{1}>0$;
$\star$=$\Omega^+_{2}>0$.} \label{fig:mu00R45}
\end{center}
\end{figure}

Here we consider a tight-binding model for ease of numerical
evaluation.  The definition of the resonant energies can be
obtained either by finding the minimum of
(\ref{2impexplicitdenom}) or from an  analysis of phase
shifts\cite{AHZdisorder}.
The solutions corresponding to each factor in Eq. (\ref{d1d2}) can
then in general be tracked as a function of separation $R$ by
minimizing ${\cal D}_{1,2}$ separately.  In practice, this works
well except in some special cases where the minima are very
shallow.  In Figure \ref{fig:mu00R45} we show the result for a
particle-hole symmetric system.  It is seen that each factor
${\cal D_\alpha}$ corresponds to an oscillating function of $R$,
with the factor determining, e.g. $\Omega_2^+$, changing from site
to site according to whether the site is even or odd.  This is due
to the strong $R$ dependence of the components $G_\alpha$; in the
simplest case, $\R \parallel (110)$ and $\mu=0$,
$G_3(\R,\omega)=G_1(\R,\omega)=0$ but $G_0(\R,\omega)\equiv
\sum_\k\, \cos (k_xR/\sqrt{2})\cos (k_yR/\sqrt{2})
\,G^0_\k(\omega)$ oscillates rapidly. At $R=0$, the problem
reduces to the double-strength single impurity case; the factor
${\cal D}_1$ gives the resonant frequency $\Omega_0^{\pm}(2V_0)$
and the factor ${\cal D}_2$ is 1. At large separation the
$\Omega_1^+$ and $\Omega_2^+$ ``envelopes" are seen to converge to
$\Omega_0^+ (V_0)$ with a length scale of a few $\xi_0 \simeq 10a$
for the parameters chosen.

\begin{figure}[tb]
\begin{center}
\leavevmode
\includegraphics[width=.9\columnwidth]{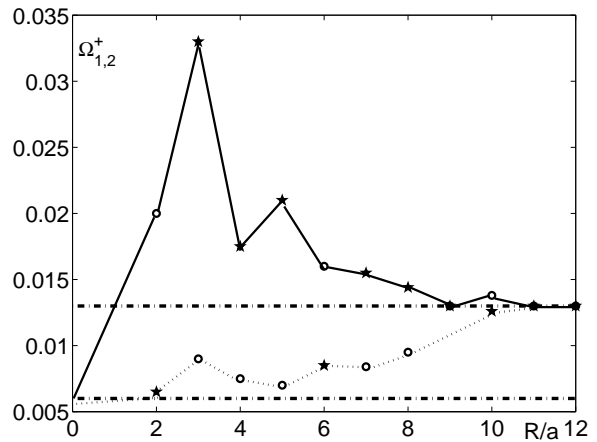}
\caption{2-impurity resonance energies $\Omega^+_{1,2}/t>0$ vs.
impurity separation $R/(a)$ for $\R\parallel (100)$, $\mu$=0,
$\Delta_0=0.1$, $V_0=10$.  Solid line: upper 2-impurity positive
resonance energy;
 dashed line: lower 2-impurity resonance energy; upper
dashed-dotted line: reference 1-impurity  resonance energy
$\Omega_0^+(V_0)$. Lower dashed-dotted line: 1-impurity resonance
energy for double impurity strength $\Omega_0^+(2V_0)$.  Symbol
indicates resonant channel: ${\circ}$=$\Omega^+_{1}>0$;
$\star$=$\Omega^+_{2}>0$.} \label{fig:mu0zerosplit1}
\end{center}
\end{figure}

In the $\R\parallel (100)$ case, the oscillations of the bound
state energies with increasing $R$ are not so simple, as seen in
Fig.\ \ref{fig:mu0zerosplit1}.  The one obvious simple difference
from the (110) case is that the energy splittings vanish much
faster with distance, as expected from the discussion in Section
\ref{subsec:gasmodel}.  Otherwise the short distance behavior of
the bound state energies is complicated.  One can check that the
energy closest to the Fermi level is $\Omega_2^+$ when $R=2+4n$,
 $n$ integer, and  $\Omega_1^+$ otherwise.

In general, the short distance behavior is difficult to analyze
analytically and we note that in neither the (110) or (100)
direction do resonances appear {\it at all} for $R=1$. Clearly the
hybridization is so strong in these cases that the picture of
perturbatively split 1-impurity states breaks down.  More
importantly, the splittings are significant out to quite large
distances.  Parameters in Figures \ref{fig:mu00R45} and
\ref{fig:mu0zerosplit1} are chosen such that $\xi_0\approx 10 a$,
as seen from Figure  \ref{fig:mu0zerosplit1}, where we indeed
expect a $e^{-R/\xi_0}$ falloff according to the previous section.
On the other hand, Figure \ref{fig:mu00R45} indicates strong
interference out to distances of 30$a$ or more! \vskip .2cm

\section{Local Density of States} \label{sec:ldos}

\subsection{General}
The spatial distribution of two-impurity bound states gives much
more detailed information about the nature of the quantum
interference processes between impurities than the bound state
energies by themselves.  Here we ask whether one can simply
express the bound state wave functions of the 2-impurity system in
terms of the 1-impurity bound states, and how they can be
classified by symmetry. We would like to make predictions for STM
experiments, including which qualitative features of the spectra
reflect the quantum numbers of these states directly, and how
these features depend on impurity potential and configuration. As
indicated in the previous section, it is important to determine
how far apart two impurities need to be to be considered
``isolated".
 Finally, we would
like to understand how robust these predictions are with respect
to changes in band structure, scattering potential, etc.

Throughout this work, the LDOS
refers to the {\em tunneling} density of states,
\begin{equation}
    \rho({\bf r},\omega) = \frac{1}{2}\sum_\sigma
    \rho_\sigma({\bfr},\omega)
    \label{eq:ldos}
\end{equation}
with the spin-resolved LDOS,
\begin{subequations}
\begin{eqnarray}
    \rho_{\uparrow}({\bf r},\omega) &=& -\pi^{-1}\mbox{Im }
    G_{11}({\bfr},\r,\omega+i0^+)\\
    \rho_{\downarrow}({\bf r},\omega) &=& +\pi^{-1}\mbox{Im }
    G_{22}({\bf r},\r,-\omega-i0^+)
\end{eqnarray}
\end{subequations}
where the subscripts 11 and 22 refer to the electron and hole
parts of the diagonal Nambu Green's function.

The factor of $1/2$ in Eq.\ (\ref{eq:ldos}) ensures that the LDOS
normalization is
\[
    \int_{-\infty}^\infty d\omega \rho(\r,\omega) = 1.
\]
Provided the peaks in the LDOS are well defined, the peak energies
agree closely with the resonance energies defined by Eq.\
(\ref{1impresenphsA}).  Some care is required, however, because
peaks in the LDOS may not appear on all sites, and can be
difficult to resolve. It is also worth noting that the peaks in
the LDOS { are not symmetric with respect to the Fermi energy},
though in practice the degree of asymmetry is very small.

 The local electron density of states measured by STM is given by
Eq. (\ref{eq:ldos}),
with \bea {\hat G}(\r,\r,\omega)&=& \sum_\k {\hat G}^0(\k,\omega)
\\&&+ \sum_{\k,\k'} e^{-i(\k-\k')\cdot \r} {\hat
G^0}(\k,\omega){\hat T}_{\k\k'}(\omega){\hat
G^0}(\k',\omega),\nn\label{STMLDOS}\eea and ${\hat T}_{\k\k'}$ is
the $T$-matrix for any number of impurities.

%
%

\subsection{Interference of 1-impurity wavefunctions.}
In the 1-impurity case, the $T$-matrix is given by  Eq.
(\ref{1imptmat}) and it is easy to see that
 \bea
- \delta G_{11}^{\,\prime\prime}(\r,\r,\omega) &=& -V_0~{\rm
Im}\left ({ (G_{11}^0(\r))^2
 \over S_-}+
{  (G_{21}^0(\r))^2  \over S_+} \right),\nn\\~\label{1impG}
 \eea
Quite generally one can express the Green's function in terms of
the exact eigenstates $\psi_n({\bf r})$ of the system in the
presence of the impurity\cite{economou}
\begin{eqnarray}
    \delta G_{11}(\r,\r, \omega)&=&\sum_n {\psi_n^* (\r)\psi_n (\r)\over \omega -\Omega_n+i0^+
    }\nn\\
    &\approx& {\psi_n^* (\r)\psi_n (\r)\over \omega -\Omega_n },
    \label{economou}
\end{eqnarray}
where the final approximation is valid for a true bound state with
$\omega$ very close to a particular bound state energy $\Omega_n$,
and will be a good approximation in the present case to the extent
the resonances are well defined, in the sense discussed above.
Comparing with the form (\ref{1impG}) thus allows us to identify
the positive and negative energy wavefunctions of the
single-impurity resonances ($V_0>0$ assumed):

\bea \psi_{\pm}(\r) = Z_\pm \left\{ \begin{array}{cc} G^0_{21}
(\r,\omega)  & ~~\omega =\Omega_0^{+}\\
G^0_{11}(\r,\omega) & ~~\omega = \Omega_0^{-}
\end{array}\right.,\label{1imppsi}
\eea where $Z_\pm$ are non-resonant wave function normalization
factors.  Note that the  electron-like bound state eigenfunction
is directly related to the off-diagonal bare Green's function,
while the hole-like wave function is proportional to the  diagonal
bare Green's function.

We can follow the same procedure for the 2 impurity Green's
function, and ask how the eigenfunctions at a particular resonant
energy are related to the single impurity wave functions we have
just found.  Since the single-impurity resonant energies are
different from the 2-impurity energies, this analysis will be
valid to the extent the splittings are small compared to
$\Omega_{0\pm}$. The Green's function $\delta G(\r,\r)$ can now be
constructed from Eq. (\ref{explictT})  and the wave functions read
off by comparing with the spectral representation in the same way
as in the 1-impurity case.  By examining (\ref{explictT}) it may
be shown that, depending on whether $\D_1$ or $\D_2$ is resonant,
the wave functions thus extracted will be of definite spatial
parity, $\psi_n(\r)=\pm \psi_n(-\r)$.  We find



\begin{eqnarray}
\psi^{p}_{+}&=&Z^p_{+} \left(G_{11p}^0+\frac{G_1({\bf
R})V_0}{{\cal D}_2^{+}}G_{12p}^0\right) ~~~\omega=\Omega_{2+}\nn\\
\psi^{p}_{-}&=&Z^p_{-} \left(G_{11p}^0-\frac{{\cal
D}_2^{-}}{G_1({\bf R})V_0}G_{12p}^0\right) ~~~\omega=\Omega_{2-}\nn\\
\psi^{s}_{+}&=&Z^s_{+}\left(G_{11s}^0-\frac{G_1({\bf
R})V_0}{{\cal D}_1^{+}}G_{12s}^0\right) ~~~\omega=\Omega_{1+}\nn\\
\psi^{s}_{-}&=&Z^s_{-} \left(G_{11s}^0+\frac{{\cal
D}_1^{-}}{G_1({\bf
R})V_0}G_{12s}^0\right)~~~\omega=\Omega_{1-}\label{2imppsi}
\end{eqnarray}
where  ${\hat G}^0_{({\scriptstyle s}, {\scriptstyle p})}\equiv
{\hat G}^0 (\r-\R/2) \pm {\hat G}^0 (\r+\R/2)$, and the
 $Z^{s,p}_\pm$ are normalization coefficients.  These are the two-impurity
odd ($p$) and even-parity ($s$) resonant state eigenfunctions
expressed directly as linear combinations of the corresponding
one-impurity eigenfunctions $\psi_\pm$ given in (\ref{1imppsi}).

\subsubsection{$\R\parallel (110)$}

We note now that in general particle and hole-like 1-impurity
eigenfunctions are mixed  in each 2-impurity state
(\ref{2imppsi}); this is possible because anomalous scattering
processes with amplitude $G_1({\bf R})$ can take place.  There are
special situations, including all configurations with $\R\parallel
(110)$, where $G_1( \R )=0$ and the eigenfunctions become much
simpler and do not mix particle and hole degrees of freedom,

\bea \psi_{\pm}^{s}(\r) = Z_\pm^s
\left\{ \begin{array}{cc} G^0_{21s}
(\r,\omega)  & ~~\omega =\Omega_{1+}\\
G^0_{11s}(\r,\omega) & ~~\omega = \Omega_{1-}
\end{array}\right.,\label{2imppsis}
\eea and \bea \psi_{\pm}^{p}(\r) = Z_\pm^p \left\{
\begin{array}{cc} G^0_{21p}
(\r,\omega)  & ~~\omega =\Omega_{2+}\\
G^0_{11p}(\r,\omega) & ~~\omega = \Omega_{2-}
\end{array}\right..\label{2imppsip}\eea
\begin{figure}[tb]
\begin{center}
\leavevmode
\includegraphics[width=1.0\columnwidth]{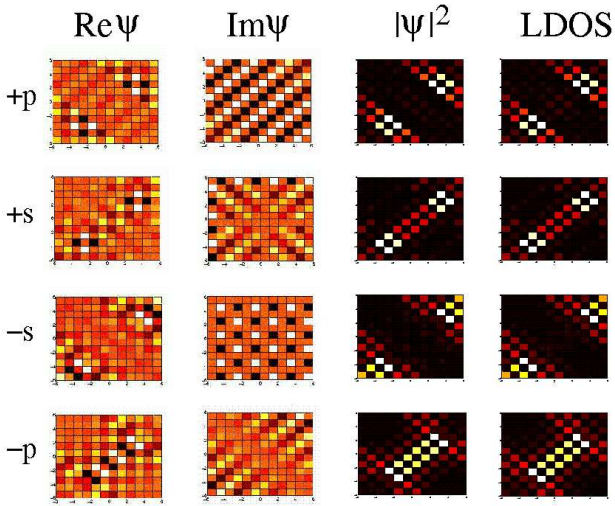}
\caption{Comparison of symmetric (s) and antisymmetric (p)
combinations of 1-particle wave functions with exact LDOS for
$\Delta_0=0.1$, $V_0=10$, $\mu=0$, $\R=(6,6)$.  2-impurity
resonance energies are $\Omega_{2\pm}/t=\pm 0.0195$,
$\Omega_{1\pm}/t=\pm 0.0075$. energies are always ordered from
highest (top) to lowest (bottom). }\label{fig:INTER}
\end{center}
\end{figure}
%

\begin{figure}[tb]
\begin{center}
\leavevmode
\includegraphics[width=.9\columnwidth]{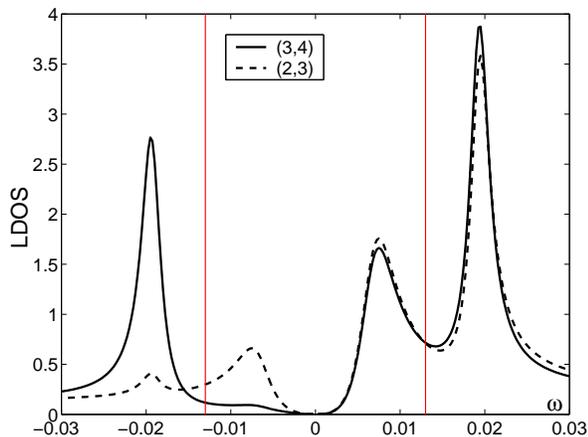}
\caption{ Comparison of LDOS spectra $\rho(\r,\omega)$ for various
$\r$ with two impurities with $\Delta_0=0.1$, $V_0$=10, $\mu=0$ at
positions $(-3,-3)$ and $(3,3)$ $(\R=(6,6))$. Vertical lines
correspond to 1-impurity resonances at $\Omega_0^\pm=\pm 0.013$.
Note difference in vertical scales between upper and lower panels.
}\label{fig:ldosvariousr}
\end{center}
\end{figure}

\begin{figure}[tb]
\begin{center}
\leavevmode
\includegraphics[width=\columnwidth]{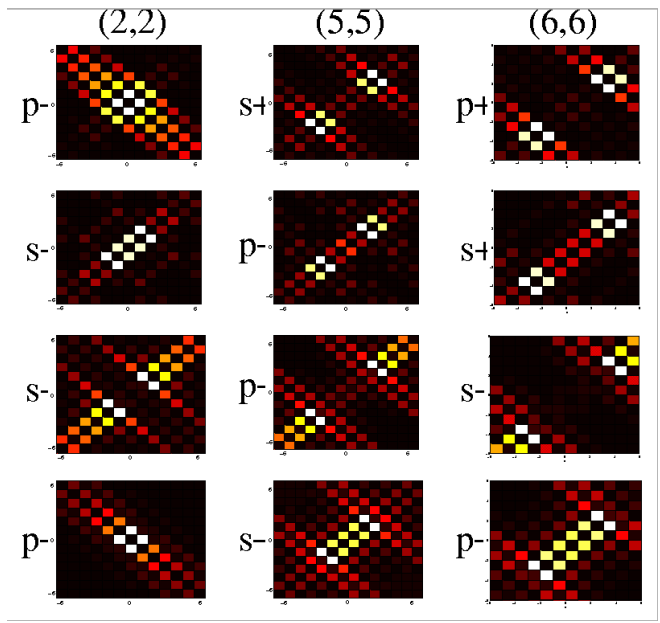}
\caption{  LDOS maps at resonant energies for ${\bf R} \parallel
(110)$. Tight binding band, $\Delta_0=0.1$, $V_0=10$, $\mu=0$
}\label{fig:LDOS110fig}
\end{center}
\end{figure}


We study this case now both to get a simple idea of how the
quantum interference between 2 impurities works in practice, and
because the (110) states are claimed to be of particular
importance for the formation of the impurity band due to the
strong interference of long range tails in the quasiparticle
wavefunctions. In Figure \ref{fig:INTER}, we have exhibited the
resonant state wavefunctions $\psi_\pm^{s,p}$ for the (110) case.
We exhibit both the wave functions themselves, so that the reader
may verify the parity of these states explicitly, and their
absolute square.    There is good agreement between the spatial
pattern of the $|\psi|^2$ as defined and the exact LDOS calculated
from (\ref{STMLDOS} ) at each resonant energy, implying that near
each resonant energy the nonresonant contributions are quite
insignificant.   It is clear that some states involve constructive
and some destructive interference between the 1-impurity
wavefunctions in different regions of space, but the spatial
patterns are, not unexpectedly, considerably more intricate than
the ``hydrogen molecule" type states one might first imagine would
form, with electrons living either directly between impurities or
completely expelled from this region.  This is of course due to
the $d$-wave character of the medium in which the quasiparticles
propagate.   For example, the LDOS is zero at the point halfway
between the two impurities for the $p$-wave states, but it is
quite small in the $s$-wave states as well. It is furthermore
clear from the Figure that both $s$ and $p$ functions can have
either constructive or destructive character, in the molecular
sense. Note that the states are shown arranged vertically
according to their eigenenergies, but  recall that the ordering of
the $s$ and $p$ ($\D_1$ and $\D_2$) states changes according to
whether $R$ is even or odd, as indicated in Figure
\ref{fig:mu00R45}.

In Figure \ref{fig:ldosvariousr}, we show the full energy
variation of the LDOS spectra on  sites near one of the impurities
to illustrate the expected widths of the resonances and the
variability with position.  Note the  surprisingly sharp
high-energy resonances, which we discuss further below. On the
other hand, on certain sites  in the configuration of Figure
\ref{fig:ldosvariousr} such as (0,0), (2,2) and (4,4), virtually
no LDOS weight at all is observed.  In Figure
\ref{fig:LDOS110fig}, we show LDOS maps for several impurity
separations along (110). Note that in the high-energy $p$ states
for ${\bf R}$=(2,2) and (6,6), both of which are fairly narrow,
nearly total destructive interference exists along the (110)
direction outside of the two impurities, but there is substantial
weight along the
(1$\overline 1$0)
direction relative to each of the two impurities.   As one moves
the impurities apart, the tails of these wavefunctions in the
(1$\overline 1$0) direction simply shift with the impurities. This
suggests that these states may be somewhat narrowed because the
quasiparticle is localized between the two impurities, but only in
the (110) direction; it may easily leak out along the (1$\overline
1$0) tails.

\subsubsection{ $\R \parallel(100)$}
\begin{figure}[tb]
\begin{center}
\leavevmode
\includegraphics[width=0.9\columnwidth]{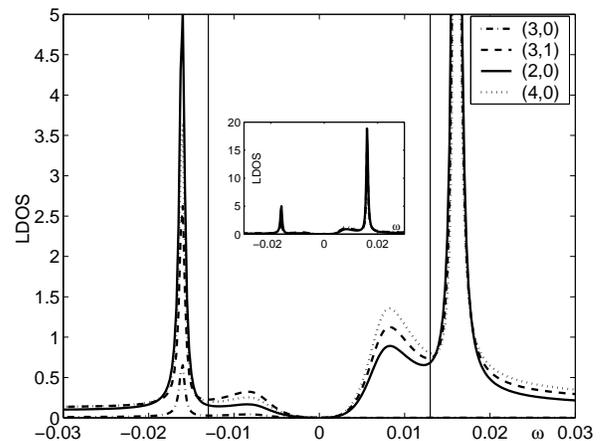}
\caption{ Comparison of LDOS spectra $\rho(\r,\omega)$ for various
$\r$ with two impurities   at positions $(-3,0)$ and $(3,0)$
$(\R=(6,0))$. Tight binding band, $\Delta_0=0.1$, $V_0=10$,
$\mu=0$.}\label{fig:ldos00variousr}
\end{center}
\end{figure}

We noted above that for large separations the intereference
between quasiparticle wavefunctions vanishes much more rapidly
with increasing separation in configurations with impurities
aligned along a crystal axis $R\parallel (100)$. For small or
intermediate spacings $R\lesssim \xi_0$, however, the bound state
splittings are just as large. Figure \ref{fig:ldos00variousr}
shows a spectrum for impurities separated by 6 lattice constants
in the (100) direction.  Although the low energy peaks are weak,
there are nevertheless four well defined peaks as expected.  The
most striking feature of Fig.\ \ref{fig:ldos00variousr} is that
the upper resonance is extremely sharp, far sharper in fact than a
single impurity resonance at the same energy!  This is
counter-intuitive based on our knowledge of the one-impurity
problem: the T-matrix denominator $S_\pm$ defined in Eq.\
(\ref{1imptmat}) has an imaginary part proportional to the density
of states of the clean $d$-wave superconductor, so that the
resonance width depends (approximately) linearly on the resonance
energy.  This is clearly not the case here.  In the two-impurity
problem where one impurity is at the origin and the second is at
$\R$, the T-matrix denominator is given by Eq.\
(\ref{2impexplicitdenom}), which can equivalently be written as
\[
    {\cal D} = \det [ V_0^{-1}\tau_3 - {\hat G}^{(1imp)}(\R,\R,\omega) ] \det {\hat T}(\omega),
\]
where ${\hat T}$ is the one-impurity $T$-matrix and ${\hat G}^{(1imp)}$ is the
Green's function with one impurity at the origin.  Thus sharp two-impurity
resonances occur for exactly the same reason as in the one-impurity case, but
because the one-impurity DOS at $\R$ is nonmonotonic in $\omega$, the resonance
broadening is not necessarily proportional to the resonance energy.

\begin{figure}[tb]
\begin{center}
\leavevmode
\includegraphics[width=\columnwidth]{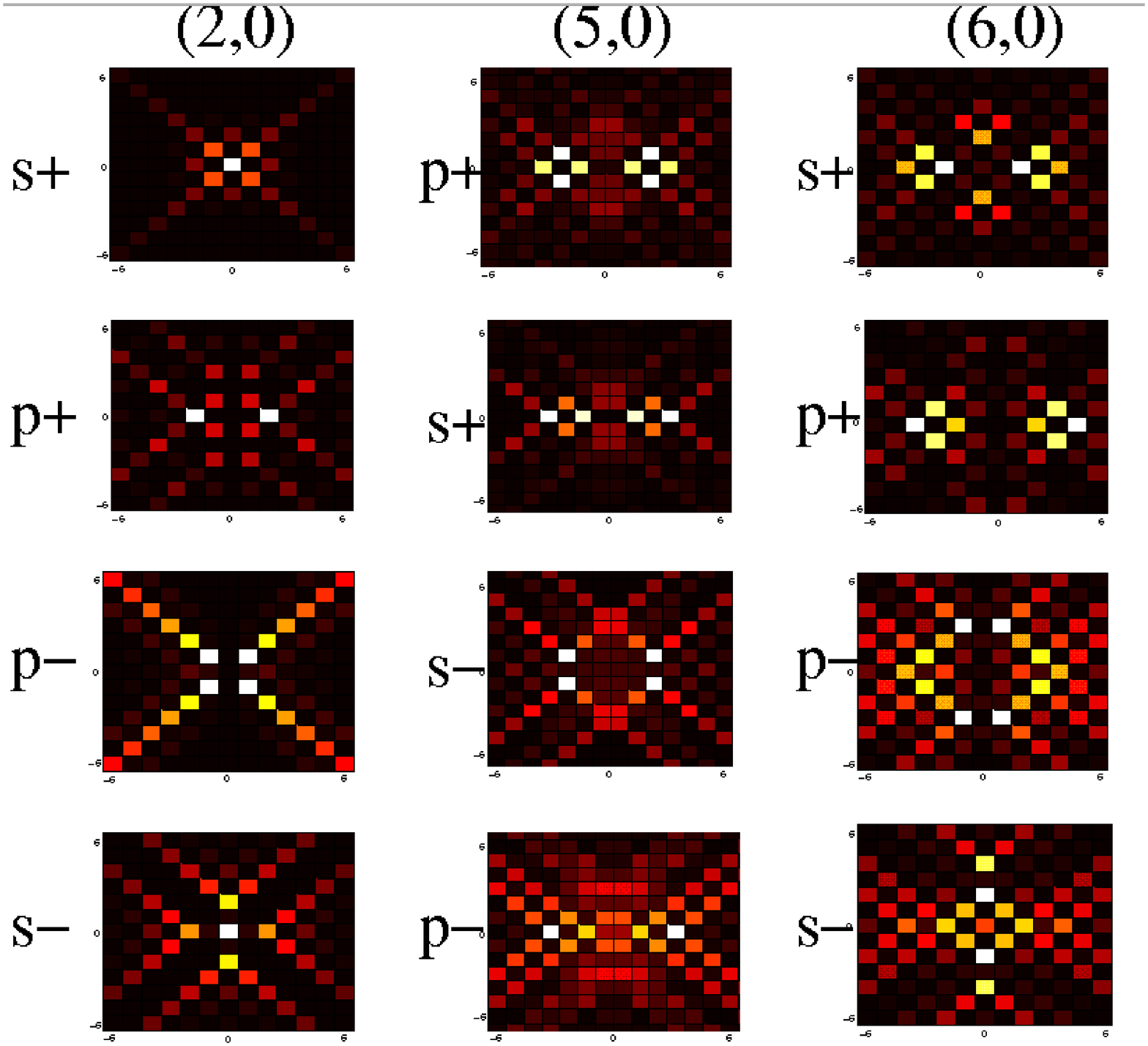}
\caption{  LDOS maps at resonant energies for $R\parallel (100)$.
Tight binding band, $\Delta_0=0.1$, $V_0=10$, $\mu=0$.
}\label{fig:LDOS100fig}
\end{center}
\end{figure}

 It seems intuitively clear that, because of destructive interference
 along the nodal directions,  2-impurity bound states
{\it could} be formed in which quasiparticles are quite
effectively trapped because they will be prevented from escaping
via the long (110) tails of the individual impurity wave
functions. In Figure \ref{fig:LDOS100fig}, the spatial structure
of the high-energy states confirms our intuition, since
quasiparticles appear to be confined primarily to the axis joining
the impurities. The 45$^\circ$ tails of these 2-impurity states
are supressed in all directions, and as expected the spectral
features are even narrower than for the ${\bf R}\parallel (110)$
configurations of Figure \ref{fig:ldosvariousr}. In the ${\bf
R}=(2,0)$ case, the quasiparticle is nearly completely confined to
the site between the two impurities, while the wave function
spreads out somewhat in the ${\bf R}=(6,0)$ case.  For these cases
$G_1({\bf R},\omega)$ vanishes as in the (110) case, so the wave
functions have the same structure as in Eq.
(\ref{2imppsis},\ref{2imppsip}). We note that when the peaks are
sharp, the resonance state is essentially localized, that is the
contribution to the wavefunction at resonance from the continuum
has vanishing weight. In general, one might have expected these
states  to decay as $1/r$ along the (110) direction and
exponentially along the (100) direction, but they may decay more
rapidly from interference effects.

Thus far we have considered only a simple tight-binding band at
half-filling, and one might worry that the existence of
quasilocalized states was a consequence of the perfect nesting of
this special electronic structure, and that such features are
unlikely to be observed in real systems.   We show below that this
is not the case, and argue that a commensuration of dominant
scattering wavevectors {\it at the bound state energy} is the
important quantity, and that finding such a state depends  on band
structure and other details.

\section{Realistic bands}
\label{sec:realistic}

\begin{figure}[tb]
\begin{center}
\leavevmode
\includegraphics[width=.9\columnwidth]{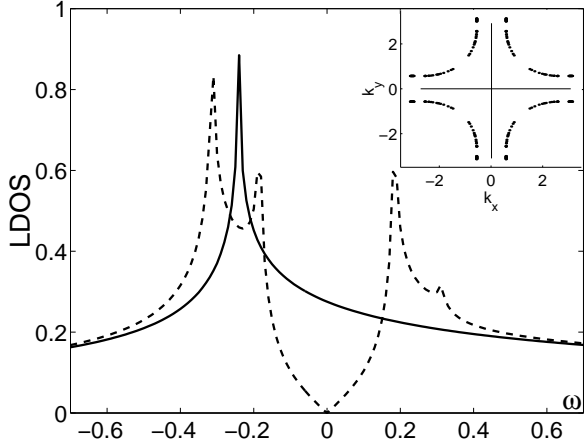}
\caption{ Solid line: BSCCO total density of states after Ref.
\protect{\cite{Norman}} vs $\omega/t_1$. Dashed line: with
$d$-wave superconducting gap of magnitude $\Delta_0=0.1t_1$.
Insert: Fermi surface at optimal doping. }\label{fig:realfermi}
\end{center}
\end{figure}
 \vskip .2cm
 \begin{figure}[tb]
\begin{center}
\leavevmode
\includegraphics[width=\columnwidth]{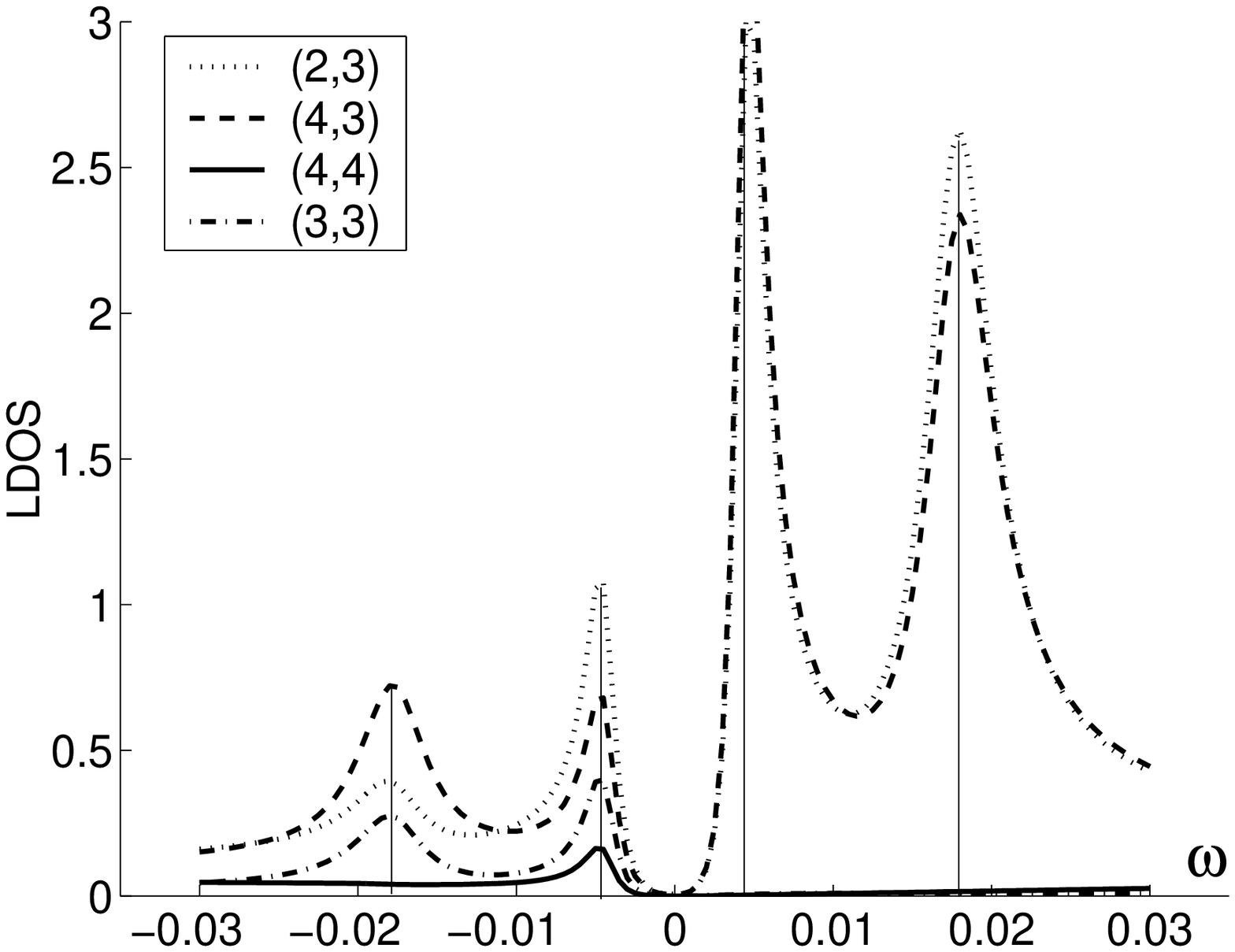}
\includegraphics[width=\columnwidth]{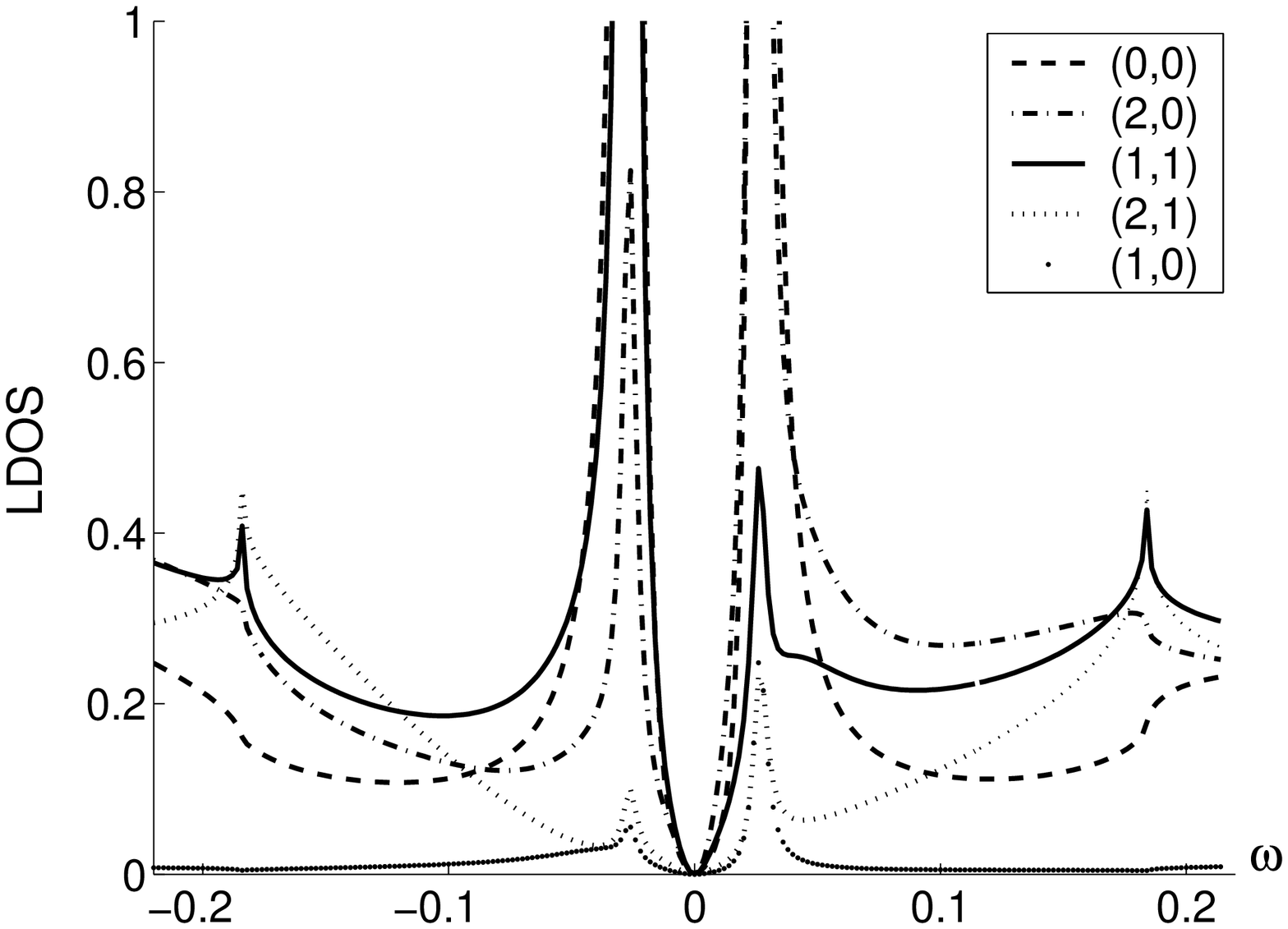}
\caption{  LDOS spectra for various sites as indicated.  Upper panel for
impurities at (-3,-3) and (3,3) (${\bf R}= (6,6)$), lower panel for impurities
at (-1,0) and (1,0) (${\bf R}= (2,0)$. Parameters for ``realistic" band (see
text), energies $\omega$ in units of $t_1$, with  $V_0=5.3$.  Note peak at
$\omega=-0.18$ in lower panel is gap edge, not impurity resonance.
}\label{fig:realisticspectra}
\end{center}
\end{figure}
Until now we have focussed on the rather artificial symmetric
tight-binding band case, both for calculational simplicity and to
illuminate the unusual density of states phenomena driven by
nesting features of the Fermi surface. Some, but clearly not all
of these phenomena will survive away for a realistic band with
particle-hole asymmetry.  Because of the intense current interest
on STM studies of the cuprates, we now focus exclusively on a
``realistic" representation of the electronic structure of
BSCCO-2212, which we parametrize by adopting the tight-binding
coefficients of Norman et al.\cite{Norman}:
\begin{eqnarray}
\epsilon(k)&=& t_0+2t_1 [\cos(k_x)+\cos(k_y)]  + 4t_2
\cos(k_x)\cos(k_y)\nonumber\\&& + 2 t_3  [\cos(2k_x)+\cos(2k_y)]
\nonumber\\&&+2t_4  [\cos(2k_x)\cos(ky)+\cos(k_x)\cos(2ky)]
\nonumber\\&&+4t_5 \cos(2kx)\cos(2ky)
\end{eqnarray}
with $t_0\dots t_5=0.879,-1,0.275,-0.087,-0.1876,0.086$ and $|t_1|\equiv 0.1488
eV$. The density of states of this band in both normal and superconducting
states is shown in Figure \ref{fig:realfermi}.

\begin{figure}[tb]
\begin{center}
\leavevmode
\includegraphics[width=\columnwidth]{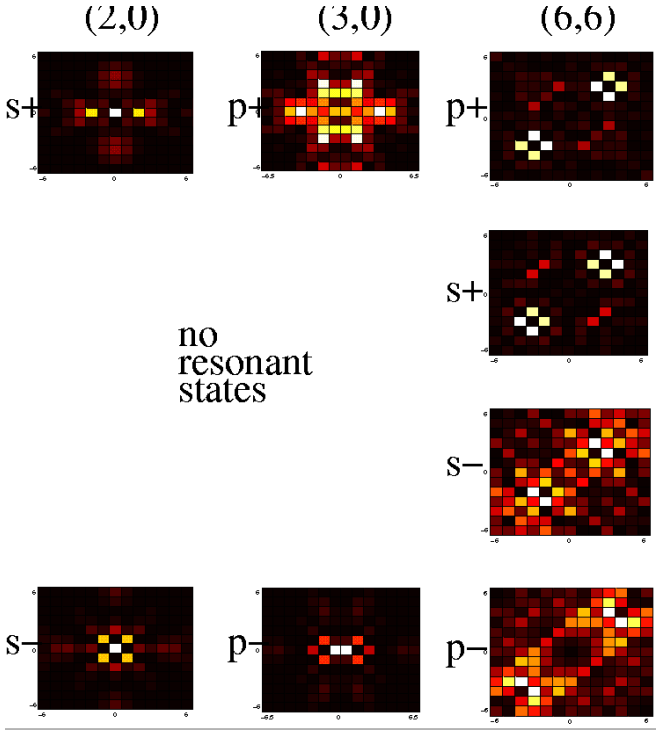}
\caption{  LDOS maps for realistic band (see text), $V_0=5.3t_1$.}
\label{fig:realLDOS}
\end{center}
\end{figure}

\subsection{Direct spectra}
We first reconsider some of the impurity configurations we treated
in Section \ref{sec:ldos} to see how the LDOS patterns and bound
states are affected by the electronic structure.  The upper panel
in  Figure \ref{fig:realisticspectra} shows the distribtion of
spectral weight on nearby sites for a configuration with
impurities at separation ${\bf R}=(6,6)$. Comparison with the
half-filled tight-binding model case shown in Figure
\ref{fig:ldosvariousr} shows little qualitative change on nearby
sites with respect to either peak positions, spectral weight or
spatial distribution. On the other hand, comparison of  LDOS maps
in Figures \ref{fig:LDOS110fig} and \ref{fig:realLDOS} show that
the longer-range character of the hole-like resonances has changed
considerably.   In other cases, particularly for small $R$, the
changes are much more drastic.  In particular, one does not
generally see all four resonances, as also found by Morr and
Stavropoulos\cite{Morr}. This is particularly true for resonances
with ${\bf R}\parallel (100)$, as illustrated in the lower panel
of Fig. \ref{fig:realisticspectra}, where the {\it low energy}
resonances have completely disappeared on all sites investigated.

\begin{figure}[tb]
\begin{center}
\leavevmode
\includegraphics[width=\columnwidth]{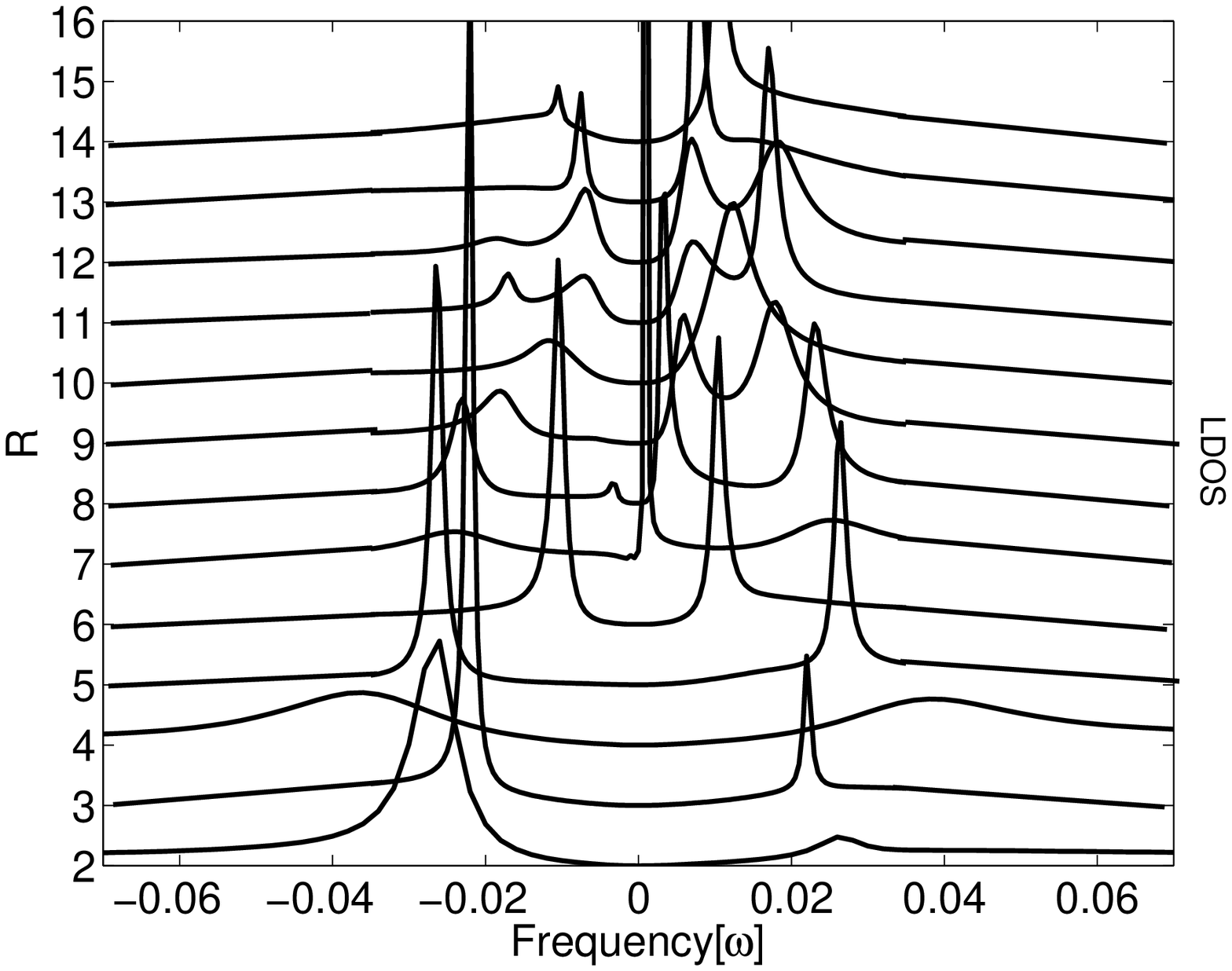}
\includegraphics[width=\columnwidth]{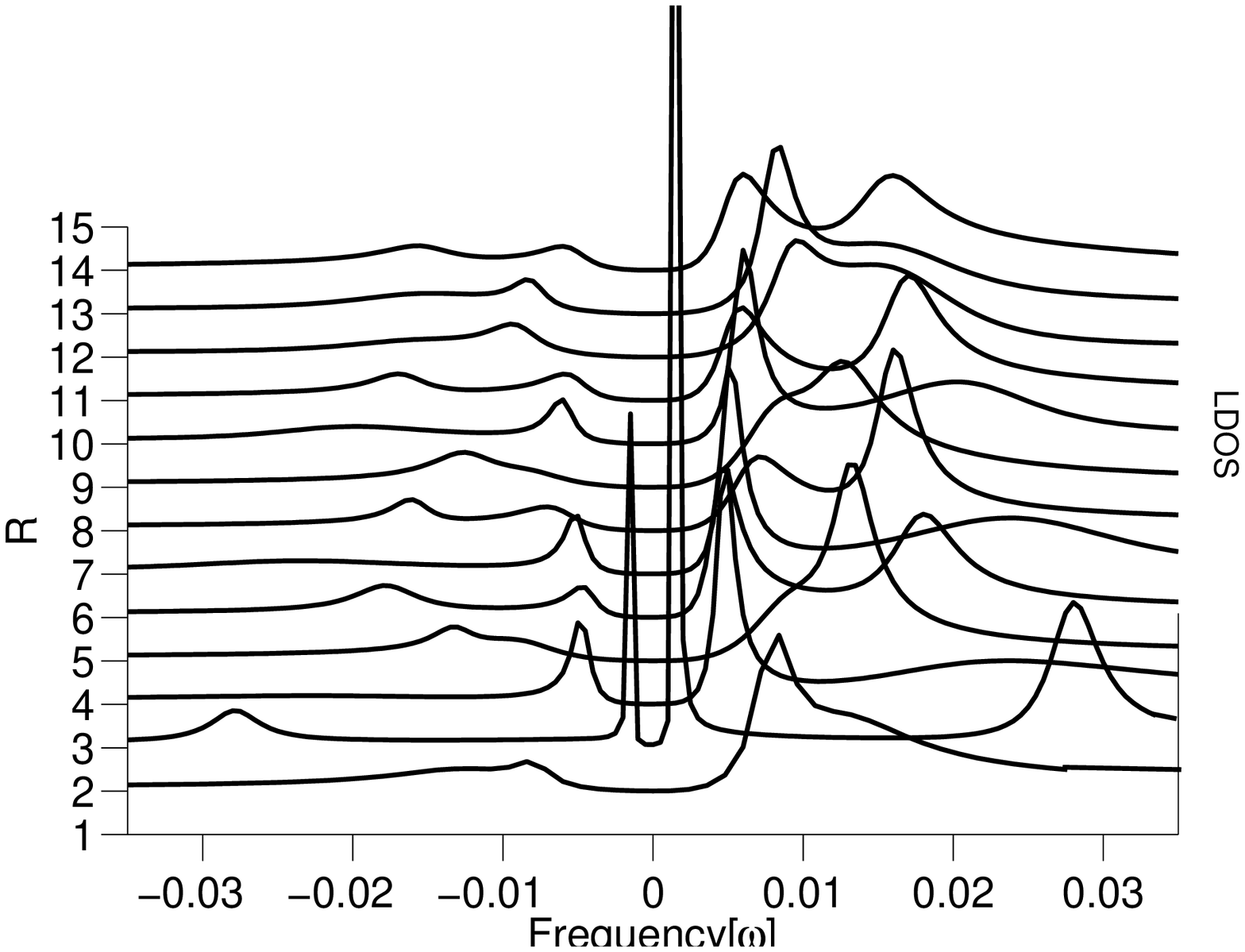}
\caption{  LDOS spectra for realistic band and $V_0=5.3t_1$ on
nearest neighbor site. Upper panel:  impurities at $(-R/2,0)$ and
$(R/2,0)$ ( ${\bf R}=(R,0)$), spectra taken at ${\bf r}=(R/2,1)$.
Lower panel: impurities at $(-R/2,-R/2)$ and $(R/2,R/2)$ ( ${\bf
R}=(R,R)$), spectra taken at ${\bf r}=(R/2,R/2+1)$.}
\label{fig:offsets}
\end{center}
\end{figure}
To give an impression of the systematics of impurity resonance
dependence on separation in the realistic system, and the
configurations in which certain resonances are overdamped, in
Figure \ref{fig:offsets} we present a series of plots of LDOS
spectra on  nearest neighbor sites along the line joining the
impurities for a range of separations with ${\bf R}||$(100) and
(110) directions.  The number of peaks in each spectrum is
variable, and for the closer separations normally only two peaks
are observed (no peaks are observed for separation 1!).   As the
impurities are moved farther apart, hybridization weakens and it
becomes easier to observe the full complement of four resonances.
  It is also clear that
the bound state splittings for the realistic band are decaying
faster in the (100) direction, but that this splitting has not
disappeared even for separations as large as $R=13$.  This
suggests that even in relatively dilute impurity systems the
assumption of isolated impurities used to analyze recent STM
experiments may need to be reexamined.

We note further that extremely sharp states occur frequently, for
both even and odd separations; clearly the commensurability
condition depends sensitively on the details of the band
structure.  Contrary to the results of Ref. \cite{Morr}, we find
that the state farthest from the Fermi level is frequently
sharpest.  In general, however, the spectra found here for the
band of Ref.\cite{Norman} are quite similar to those obtained in
Ref.\cite{Morr} when direct comparisons are possible.

The condition for a true bound state [see  Eq.\
(\ref{2impexplicitdenom})] is satisfied at frequency $\omega$ by
$\hat G^0(\R,\omega) \hat T \hat G^0(\R,\omega) \hat T = 1$. Since
this must be satisfied independently for real and imaginary parts
of $G^0(\R,\omega)$, sharp resonances only appear for selected
impurity separations.  The product $\hat G^0(\R,\omega) \hat T
\hat G^0(\R,\omega) \hat T$ in Eq.\ (\ref{2impexplicitdenom}) is
equivalently written
\begin{equation}
    \sum_{\k,\qq} e^{i\qq\cdot \R} \hat G(\k,\omega) \hat T(\omega)
        \hat G(\k + \qq ,\omega) \hat T(\omega).
    \label{jointdos}
\end{equation}
It was argued by Hoffman et al.\cite{Hoffman2}, in an analysis of
disordered BSCCO samples, that the characteristic wavevectors
found in the spatial Fourier transform of the LDOS are determined
by peaks in the joint density of states $\sum_\k \mbox{Im }
G_{11}(\k,\omega) \mbox{Im } G_{11}(\k+\qq,\omega)$.  It is
interesting to ask whether these same $q$-vectors are seen in Eq.\
(\ref{jointdos}).
\begin{figure}[tb]
\begin{center}
\includegraphics[width=.9\columnwidth]{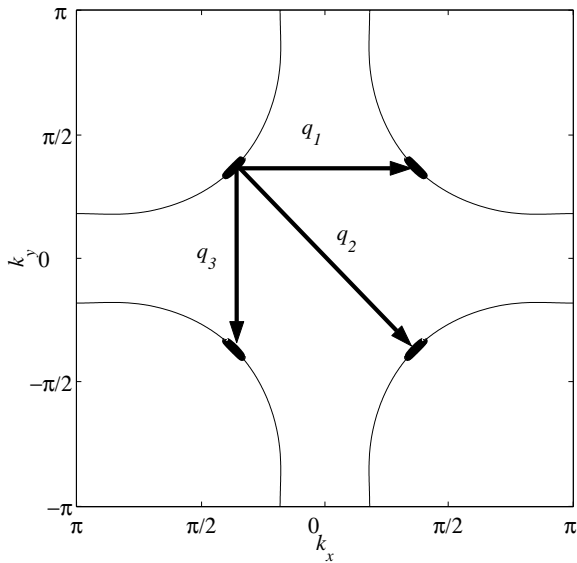}
\caption{  Fermi surface of $BSCCO-2212$ with constant energy
surfaces at $\omega = 0.04$ shown as small filled ellipses at the
nodal points.  ${\bf q_1}$, ${\bf q_2}$, ${\bf q_3}$ are wave
vectors for which the joint density of states is large.}
\label{fig:banana}
\end{center}
\end{figure}

At the small energies considered here, 3 distinct $q$-vectors
contribute to the joint DOS [a fourth, $q=0$ does not produce any
oscillatory $\R$-dependence in Eq.\ (\ref{jointdos})], as
illustrated in Fig.\ \ref{fig:banana} (we neglect, for states
sufficiently close to unitarity, the distinction between the tips
of the quasiparticle constant energy contours and the nodal wave
vectors). As a simple example, consider
the realistic band model in the (100) direction, shown in the
upper panel of Figure \ref{fig:offsets}.
The $\R$-dependence of Eq.\ (\ref{jointdos}) is straightforward
since $\qq_1\cdot (1,0) = \qq_2\cdot (1,0) \approx 2.28/a$, and
$\qq_3 \cdot \R = 0$, where $a$ is the lattice constant.  Naively,
the standing-wave condition for a particle trapped between the two
impurities is $(\qq_1\cdot \R + 2\eta_0) = n\pi$, where $\eta_0 =
\tan^{-1} (\pi N_0V_0)$ is the one-impurity scattering phase
shift. On the other hand, for $\R$ in the (110) direction, we must
simultaneously satisfy the commensurability requirements
$(\qq_1\cdot \R + 2\eta_0) = (\qq_3\cdot \R + 2\eta_0) = n\pi$,
and $(\qq_2\cdot \R + 2\eta_0) = m\pi$ to form a resonant state,
and we see that---as we observe in Fig.\ \ref{fig:offsets}---sharp
resonances occur much less frequently in the (110) than in the
(100) direction. Quantitatively, the criterion for standing wave
formation is approximately satisfied for $\R = (3,0)$, $(7,0)$,
$(11,0)$, $(14,0)$ and $\R = (3,3)$, $(11,11)$, which generally
agrees with Fig.\ \ref{fig:offsets}.


We stress, however, that the relative success of this naive
picture at making quantitative predictions is a bit surprising.
There is nothing in our consideration to account for the
$\omega$-dependence of the $T$-matrix, or for the Nambu structure
of the Green's functions. Furthermore, the approximation of the
integral over $\qq$ in Eq.\ (\ref{jointdos}) by a sum over a few
dominant wavevectors is not expected to by justified at a
quantitative level.  Nonetheless, our considerations seem to
indicate that the long-lived two-impurity bound states are derived
from a few selected wavevectors.

\begin{figure}[tb]
\begin{center}
\leavevmode
\includegraphics[width=0.9\columnwidth]{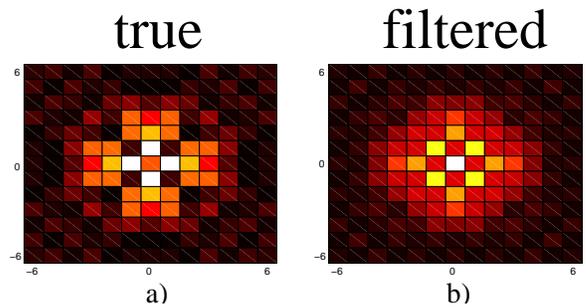}
\caption{  LDOS map for 1 strong repulsive impurity ($V_0=5.3$) at
hole type resonance $\Omega_0^-=0.011t_1$ in system with
``realistic band" (see text) a) without and b) with filter.
}\label{fig:1impfilter}
\end{center}
\end{figure}
\subsection{Filtering effects}
The discrepancies  between the simple picture of a Zn impurity as
a strong potential scatterer in a $d$-wave superconducting host
and the LDOS measured near $Zn$ impurities in BSCCO-2212 samples
have been alluded to above.  Martin and
Balatsky\cite{martinbalatsky} proposed that this problem could be
resolved by noting that electrons must first tunnel through the
BiO layer before reaching the CuO$_2$ plane; applying the
appropriate matrix elements for this process led them to a picture
of a ``filtered" DOS in which, in the simplest version, the STM
tip samples not the LDOS corresponding to the atom directly under
it, but rather to a sum of the LDOS on the surrounding 4 nearest
neighbor sites.  With this ansatz, the LDOS pattern surrounding a
Zn atom becomes, at a resonant energy of -1.5meV, rather similar
to the experimentally observed one, with a bright spot at the
center of the pattern, see Figure \ref{fig:1impfilter}.
\begin{figure}[tb]
\begin{center}
\leavevmode
\includegraphics[width=\columnwidth]{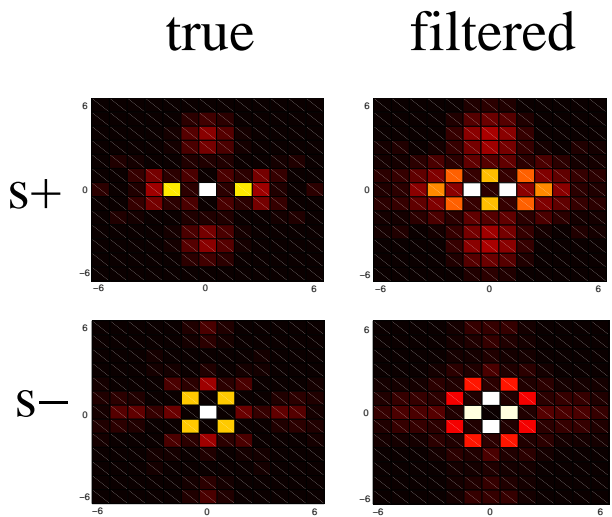}

\caption{  Comparison of true and filtered LDOS for two impurities at (-1,0)
and (1,0) ${\bf R}=(2,0)$. $V_0=5.3$, $|\Omega_1^\pm|=0.026t_1$, ``realistic
band" (see text).} \label{fig:comparefilt}
\end{center}
\end{figure}

We now point out the rather obvious fact that this filtering
mechanism is characteristic of the presence of the BiO layer in
the BSCCO-2212 system, and should therefore be present in any STM
measurement.  If two nearby impurities are located via their
resonant signals in such a measurement, the same filter should be
applied to extract the true LDOS of the superconducting CuO$_2$
layer. {\it If} it is found that the filtering mechanism works
only in the case of isolated impurities, but the observed pattern
in the 2-impurity case is quite different than that predicted by
the simple filtered potential model, it must be abandoned and more
sophisticated explanations sought.  For example, it will be
interesting to pursue the alternative ``Kondo" explanation of
Polkovnikov et al.\cite{vojta} in the case of the 2-impurity
problem. If one takes this model seriously, bringing two
impurities close together should induce an RKKY interaction
between the local moments on each impurity site, supressing the
local Kondo screening and thereby weakening each impurity's
scattering phase shift.  One might then naively expect, in such a
scenario, that bound state energies would generally be found at
higher energies than in the isolated impurity case.  Of course,
one's intuition based on the 2-Kondo impurity problem in a normal
metal is to be distrusted in this case, where the linear bare
density of states already makes the 1-impurity problem
critical\cite{Ingersent}.

\section{Conclusions}

In this paper we have explored a number of aspects of the quantum interference
of impurity bound states in $d$-wave superconductors. We gave the exact form of
the 2-impurity $t$-matrix for two potentials separated by ${\bf R}$, and showed
that in general it has four resonances at frequencies $\pm \Omega_1$ and $\pm
\Omega_2$ which depend on ${\bf R}$. In simple situations, the eigenfunctions
of the 2-impurity resonant states can be constructed explicitly in terms of the
eigenfunctions of the 1-impurity problem.  Depending on the impurity
configuration and electronic structure, some of the resonances are overdamped
on specific sites or indeed sometimes over the entire lattice, leading to a
smaller number of visible resonances  in special situations.  On the other
hand, in other  situations resonant states were observed in the 2-impurity
problem which were much sharper than their 1-impurity counterparts, in some
cases occuring quite far from the Fermi level, contradicting one's intuition
that these states should be more strongly damped.  We have interepreted these
states as impurity "traps" in which quasiparticles are hindered  by quantum
interference, over surprisingly long lifetimes, from leaking out of the region
between the two impurities.

 The splitting of the bound state energies relative to
the 1-impurity case was studied, and it was shown that the parity and energy of
the 2-impurity eigenfunctions  oscillate as a function of impurity separation.
At asymptotically large distances the splittings were shown to vary as $\sim
1/{R}$ for impurities aligned along the (110) direction and $\sim \exp
(-R)/\sqrt{R}$ along (100). Systematic STM measurements of these splittings for
isolated pairs of impurities at different ${\bf R}$ were shown to provide a
direct measurement of the spatial dependence of the Green's functions of the
bulk superconductor.

Finally we calculated the local density of states for 2 impurities
in a realistic band characteristic of the BSCCO-2212 system on
which most STM experiments have been performed.  The one
qualitative difference relative to the particle-hole symmetric
case we examined earlier was the overdamping of some bound states
on {\it any} lattice site we studied for certain impurity
configurations.  This makes it clear that  even some qualitative
features of LDOS spectra with two impurities will depend on
details of the system in question.  To extract information from
STM when 2-impurity configurations are isolated will therefore
require a careful fit to theory.  We have made predictions for
several concrete situations which can be tested if such
configurations can be found.  In particular, we have calculated
the density of states for realistic parameters corresponding to a
Zn impurity in BSCCO, and given results for both the direct LDOS
and for the ``filtered" LDOS proposed by Martin and
Balatsky\cite{martinbalatsky} to explain discrepancies in the
standard model of Zn as a potential scatterer when compared with
experiment.  If the ``filter" works for isolated single Zn
impurities but not for pairs of Zn atoms, it would be strong
evidence in favor of an explanation for the Zn results in terms of
residual induced local magnetism of the defect.

We close by remarking that the solution of the 2-impurity problem
may have important implications for the disordered N-impurity
$d$-wave superconductor and the interpretation of STM experiments.
In particular, we have shown on the one hand that pairs of
impurities can give rise to trapped states which have great deal
of spectral weight; on the other hand, interference from other
impurities can destroy the characteristic pattern expected for an
``isolated" impurity even when they are widely separated. A more
thorough investigation of these questions requires a careful
comparison with the many-impurity system.  One hint of the
importance of the 2-impurity states in the fully disordered system
comes from the study of the perfectly nested band, where we find
that many of the unusual symmetry-based features of the total
density of states\cite{yashenkin}
 are reflected already in the simple 2-impurity problem as well.
In addition, we have investigated the effects of self-consistent
treatment of the order parameter on the results above, which were
all produced assuming homogeneous $\Delta_\k$. We find that,
although spectral weight is shifted by the order parameter
supression around the impurity site, in general away from the
Fermi level\cite{atkinsonops}, the LDOS patterns are rather weakly
affected. We will report in detail on these findings
elsewhere\cite{AHZdisorder}.


\vskip .2cm
 {\it
Acknowledgements} This work was partially supported by NSF grant
NSF-DMR-9974396, BMBF 13N6918/1 and  the Alexander von Humboldt Foundation. PH
and LZ would also like to thank J. Mannhart and Lehrstuhl Experimentalphysik VI
of the University of Augsburg for hospitality during preparation of the
manuscript.

\end{document}